\def\year{2023}\relax
\documentclass[letterpaper]{article} % DO NOT CHANGE THIS
\usepackage{aaai22}  % DO NOT CHANGE THIS
\usepackage{times}  % DO NOT CHANGE THIS
\usepackage{helvet}  % DO NOT CHANGE THIS
\usepackage{courier}  % DO NOT CHANGE THIS
\usepackage[hyphens]{url}  % DO NOT CHANGE THIS
\usepackage{graphicx} % DO NOT CHANGE THIS
\usepackage{natbib}  % DO NOT CHANGE THIS AND DO NOT ADD ANY OPTIONS TO IT
\usepackage{caption} % DO NOT CHANGE THIS AND DO NOT ADD ANY OPTIONS TO IT
\DeclareCaptionStyle{ruled}{labelfont=normalfont,labelsep=colon,strut=off} % DO NOT CHANGE THIS
\usepackage{algorithm}
\usepackage{algorithmic}

\usepackage{newfloat}
\usepackage{listings}
\floatstyle{ruled}
\newfloat{listing}{tb}{lst}{}
\floatname{listing}{Listing}
\usepackage{textcomp}
\usepackage{xcolor}
\usepackage{listings}
\usepackage{indentfirst}
\usepackage{amsfonts}
\usepackage{graphicx}
\usepackage{graphbox}
\usepackage{amsmath}
\usepackage{amsthm}
\usepackage{cases}

\newtheorem{theorem}{Theorem}

\newtheorem{lemma}{Lemma}
\newtheorem{definition}{Definition}

\newtheorem{proposition}{Proposition}

\usepackage{subfigure} 
\usepackage[section]{placeins}
\usepackage{float}
\usepackage{multirow}

\newcommand*{\Scale}[2][4]{\scalebox{#1}{$#2$}}%

\title{When Congestion Games Meet Mobile Crowdsourcing: Selective Information Disclosure}

\author{
    Hongbo Li, Lingjie Duan
}
\affiliations{
    Singapore University of Technology and Design\\
    hongbo\_li@mymail.sutd.edu.sg, lingjie\_duan@sutd.edu.sg
}

\begin{document}

\captionsetup[figure]{labelfont={default},labelformat={default},labelsep=period,name={Fig.}} 

\maketitle

\begin{abstract}
In congestion games, users make myopic routing decisions to jam each other, and the social planner with the full information designs mechanisms on information or payment side to regulate. However, it is difficult to obtain time-varying traffic conditions, and emerging crowdsourcing platforms (e.g., Waze and Google Maps) provide a convenient way for mobile users travelling on the paths to learn and share the traffic conditions over time. When congestion games meet mobile crowdsourcing, it is critical to incentive selfish users to change their myopic routing policy and reach the best exploitation-exploration trade-off. By considering a simple but fundamental parallel routing network with one deterministic path and multiple stochastic paths for atomic users, we prove that the myopic routing policy's price of anarchy (PoA) is larger than $\frac{1}{1-\rho}$, which can be arbitrarily large as discount factor $\rho\rightarrow1$. To remedy such huge efficiency loss, we propose a selective information disclosure (SID) mechanism: we only reveal the latest traffic information to users when they intend to over-explore the stochastic paths, while hiding such information when they want to under-explore. We prove that our mechanism reduces PoA to be less than $\frac{1}{1-\frac{\rho}{2}}$. Besides the worst-case performance, we further examine our mechanism's average-case performance by using extensive simulations.
\end{abstract}

\section{Introduction}
In transportation networks of limited bandwidth, mobile users are selfish to choose routing decisions myopically and aim to minimize their own travel costs on the way. Traditional congestion games study such selfish routing to understand the efficiency loss using the concept of the price of anarchy (PoA) \cite{roughgarden2002bad,cominetti2019price,bilo2020price,hao2022price}. To regulate atomic or non-atomic users' selfish routing and reduce social cost, various incentive mechanisms are designed by using  monetary payments to penalize users travelling on undesired paths \cite{brown2017optimal,ferguson2021effectiveness,li2022online}. As it may be difficult to implement such payments on users, non-monetary mechanisms are also designed to provide information restriction on selfish users to change their routing decisions to approach the social optimum \cite{tavafoghi2017informational,sekar2019uncertainty,castiglioni2021signaling}.
However, these works largely assume that the social planner has full information of all traffic conditions, and limit attentions to an one-shot static scenario to regulate. 

In common practice, the traffic information dynamically changes over time and is difficult to predict in advance \cite{nikolova2011stochastic}. To obtain such time-varying information, emerging traffic navigation platforms (e.g., Waze and Google Maps) crowdsource mobile users to learn and share their observed traffic conditions on the way \cite{vasserman2015implementing,zhang2018distributed}. However, such platforms make all information public, and current users still make selfish routing decisions to the path with shortest travel latency, instead of choosing diverse paths to learn more information for future users. As a stochastic path's traffic condition alternates between congestion states over time, the platforms may miss enough exploration to reduce the social cost. 

There are some recent works studying information sharing among users in a dynamic scenario. For example, \citeauthor{meigs2017learning} (\citeyear{meigs2017learning}) and \citeauthor{wu2019learning} (\citeyear{wu2019learning}) make use of former users' observation to help learn the future travel latency and converge to the Wardrop Equilibrium under full information. Similarly, \citeauthor{vu2021fast} (\citeyear{vu2021fast}) design an adaptive information learning framework to accelerate convergence rates to Wardrop equilibrium for stochastic congestion games. However, these works cater to users' selfish interests and do not consider mechanism design to motivate users to reach social optimum. To study the social cost minimization, multi-armed bandit (MAB) problems are also formulated to derive the optimal exploitation-exploration policy among multiple stochastic arms (paths) \cite{gittins2011multi,krishnasamy2021learning}. Recently, \citeauthor{bozorgchenani2021computation} (\citeyear{bozorgchenani2021computation}) apply MAB models to predict the network congestion in a fast changing vehicular environment. However, all of these MAB works strongly assume that users upon arrival always follow the social planner’s recommendations and overlook users’ deviation to selfish routing. 

When congestion games meet mobile crowdsourcing, how to incentive selfish users to listen to the social planner's optimal recommendations is our key question in this paper. As traffic navigation platforms seldom charge users, we target at non-monetary mechanism design which satisfies budget balance in nature. Yet we cannot borrow those information mechanisms from the literature in mobile crowdsourcing, as their considered traffic information is exogenous and does not depend on users' routing decisions \cite{kremer2014implementing,papanastasiou2018crowdsourcing,li2017dynamic,li2019recommending}. For example, \citeauthor{li2019recommending} (\citeyear{li2019recommending}) consider a simple two-path transportation network, one with deterministic travel cost and the other alternates over time between a high and a low stochastic cost states due to external weather conditions. In their finding, a selfish user is always found to under-explore the stochastic path to learn latest information there for future users. In our congestion problem, however, a user will add himself to the traffic flow and change the congestion information in the loop. Thus, we imagine users may not only under-explore but also over-explore stochastic paths over time.
Furthermore, since the congestion information (though random) depends on users' routing decisions, it is easier for a user to reverse-engineer the system states based on the platform's optimal recommendation. In consequence, the prior information hiding mechanisms \cite{tavafoghi2017informational,li2019recommending,zhu2022information} become no longer efficient. 

We summarize our key novelty and main contributions in this paper as follows.
\begin{itemize}
    \item \emph{Mechanism design when congestion games meet mobile crowdsourcing:} To our best knowledge, this paper is the first to regulate atomic users' routing over time to reach the best exploitation-exploration trade-off by providing incentives. In Section \ref{section2}, we model a dynamic congestion game in a transportation network of one deterministic path and multiple stochastic paths to learn by users themselves. When congestion games meet mobile crowdsourcing, our study extends the traditional congestion games fundamentally to create positive information learning generated by users themselves.
    \item \emph{POMDP formulation and PoA analysis:} In Section \ref{section3}, we formulate users' dynamic routing problems using the partially observable Markov decision process (POMDP) according to hazard beliefs of risky paths. 
    Then in Section \ref{section4}, we analyze both myopic and socially optimal policies to learn stochastic paths' states, and prove that the myopic policy misses both exploration (when strong hazard belief) and exploitation (when weak hazard belief) as compared to the social optimum. Accordingly, we prove that the resultant price of anarchy (PoA) is larger than $\frac{1}{1-\rho}$, which can be arbitrarily large as discount factor $\rho\rightarrow1$.
    \item \emph{Selective information disclosure (SID) mechanism to remedy efficiency loss:} In Section \ref{section5}, we first prove that the prior information hiding mechanism in congestion games makes PoA infinite in our problem. Alternatively, we propose a selective information disclosure mechanism: we only reveal the latest traffic information to users when they over-explore the stochastic paths, while hiding such information when they under-explore. We prove that our mechanism reduces PoA to be less than $\frac{1}{1-\frac{\rho}{2}}$, which is no larger than $2$. Besides the worst-case performance, we further examine our mechanism's average-case performance by using extensive simulations.
\end{itemize}

We provide our simulation code here. \footnote{https://github.com/redglassli/Congestion-games-SID}

\begin{figure}[t]
    \centering
    \subfigure[A typical parallel transportation network with $N+1$ paths.]{
    \includegraphics[width=0.43\textwidth]{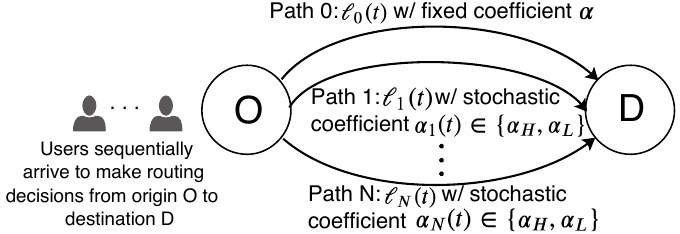}\label{fig:congestion_game}}
    \hspace{1cm}
    \subfigure[The partially observable Markov chain for modelling $\alpha_i(t)$ dynamics of stochastic path $i\in\{1,\cdots,N\}$.]{
    \includegraphics[width=0.43\textwidth]{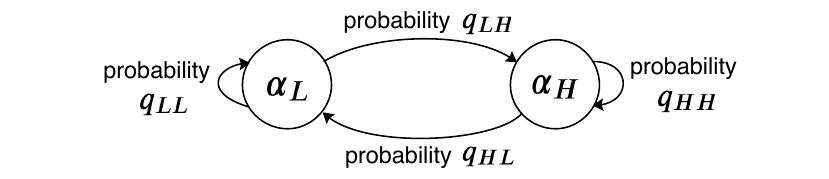}\label{fig:POMDP}}
    \captionsetup{font={footnotesize}} 
    \caption{At the beginning of each time slot $t\in\{1,2,\cdots\}$, a user arrives to choose a path among $N+1$ paths in the transportation network in Fig. \ref{fig:congestion_game}. The current travel latency $\ell_i(t)$ of each path $i\in\{0,1 ...,N\}$ has linear correlation with last latency $\ell_i(t-1)$ and evolves according to current user choice in (\ref{L_0(t+1)}) and (\ref{L_i(t+1)}). Path 0 is a safe route and its latency has fixed correlation coefficient $\alpha\in(0,1)$ to change from last round. Yet any risky path $i\in\{1,\cdots,N\}$ has a stochastic correlation coefficient $\alpha_i(t)$, which alternates between low coefficient state $\alpha_L\in[0,1)$ and high state $\alpha_H\geq 1$ according to the partially observable Markov chain in Fig. \ref{fig:POMDP}.}
    \label{pricing_fig}
\end{figure}

\section{System Model}\label{section2}
As illustrated in Fig. \ref{fig:congestion_game}, we consider a dynamic congestion game lasting for infinite discrete time horizon. At the beginning of each time epoch $t\in\{1, 2, \cdots\}$, an atomic user arrives to travel on one out of $N+1$ paths from origin O to destination D. Similar to the existing literature of congestion games (e.g., \citeauthor{kremer2014implementing} \citeyear{kremer2014implementing}; \citeauthor{tavafoghi2017informational} \citeyear{tavafoghi2017informational}; \citeauthor{li2019recommending} \citeyear{li2019recommending}), in Fig. \ref{fig:congestion_game} the top path 0 as a safe route has a fixed traffic condition $\alpha$ that is known to the public, while the other $N$ bottom paths are risky/stochastic to alternate between traffic conditions $\alpha_L$ and $\alpha_H$ over time. Thus, the crowdsourcing platform expects users to travel to risky paths from time to time to learn the actual traffic information and plan better routing advisory for future users.

In the following, we first introduce the dynamic congestion model for the transportation network, and then introduce the users' information learning and sharing in the crowdsourcing platform. 

\subsection{Dynamic Congestion Model}
Let $\ell_i(t)$ denote the travel latency of path $i\in \{0,1,\cdots,N\}$ estimated by a new user arrival on path $i$ at the beginning of each time slot $t\in\{1,2,\cdots\}$. Then the current user decides the best path $i\in\{0,1,\cdots,N\}$ to choose by comparing the travel latencies among all paths. We denote a user's routing choice at time $t$ as $\pi(t)\in \{0,1,\cdots,N\}$. 
For this user, he predicts $\ell_i(t)$ based on the latest latency $\ell_i(t-1)$ and the last user's decision $\pi(t-1)$. 

Some existing literature of delay pattern estimation (e.g., \citeauthor{ban2009delay} \citeyear{ban2009delay}; \citeauthor{alam2019prediction} \citeyear{alam2019prediction}) assumes that $\ell_i(t+1)$ is linearly dependent on $\ell_i(t)$. Thus, for safe path 0 with the fixed traffic condition, its next travel latency $\ell_0(t+1)$ changes from $\ell_0(t)$ with constant correlation coefficient $\alpha$. Here $\alpha\in(0,1)$ measures the leftover flow to be serviced over time. Yet, if the current atomic user chooses this path (i.e., $\pi(t)=0$), he will introduce an addition $\Delta \ell$ to the next travel latency $\ell_0(t+1)$, i.e., 
\begin{equation}
    \ell_0(t+1)=\begin{cases}
    \alpha \ell_0(t)+\Delta \ell,&\text{if }\pi(t)=0,\\
    \alpha \ell_0(t),&\text{if }\pi(t)\neq 0.
    \end{cases}\label{L_0(t+1)}
\end{equation}
Differently, on any risky path $i\in\{1,\cdots,N\}$, its correlation coefficient $\alpha_i(t)$ in this round is stochastic due to the random traffic condition (e.g., accident and weather change) at each time slot $t$. Similar to the congestion game literature \cite{meigs2017learning}, we suppose $\alpha_i(t)$ alternates between low coefficient state $\alpha_L\in[0,1)$ and high state $\alpha_H\in[1,+\infty)$ below:
\begin{equation*}
    \alpha_i(t)=\begin{cases}
    \alpha_L, &\text{if path } i\text{ has a good traffic condition at $t$,}\\
    \alpha_H, &\text{if path } i\text{ has a bad traffic condition at $t$.}
    \end{cases}
\end{equation*}
Note that we consider $\alpha_L<\alpha<\alpha_H$ such that each path can be chosen by users and we also allow jamming on risky paths with $\alpha_H\geq 1$. The transition of $\alpha_i(t)$ over time is modeled as the partially observable Markov chain in Fig.~\ref{fig:POMDP}, where the self-transition probabilities are $q_{LL}$ and $q_{HH}$ with $q_{LL}+q_{LH}=1$ and $q_{HH}+q_{HL}=1$. Then the travel latency $\ell_i(t+1)$ of any risky path $i\in\{1,\cdots,N\}$ is estimated as
\begin{equation}
    \ell_i(t+1)=\begin{cases}
    \alpha_i(t) \ell_i(t)+\Delta \ell,&\text{if }\pi(t)=i,\\
    \alpha_i(t) \ell_i(t),&\text{if }\pi(t)\neq i.
    \end{cases}\label{L_i(t+1)}
\end{equation}
To obtain this $\alpha_i(t)$ realization for better estimating future $\ell_i(t+1)$ in (\ref{L_i(t+1)}), the platform may expect current user to travel on this risky path $i$ to learn and share his observation.

\subsection{Crowdsourcing Model for Learning}
After choosing a risky path $i\in\{1,\cdots,N\}$ to travel, in practice a user may not obtain the whole path information when making local observation and reporting to the crowdsourcing platform. Two different users travelling on the same path may have different experiences. Similar to \citeauthor{li2019recommending} (\citeyear{li2019recommending}), we model $\alpha_i(t)$ dynamics as the partially observable two-state Markov chain in Fig.~\ref{fig:POMDP} from the user point of view. We define a random observation set $\mathbf{y}(t)=\{y_1(t),\cdots,y_N(t)\}$ for $N$ risky paths, where $y_i(t)\in\{0,1,\emptyset\}$ denotes the traffic condition of path $i$ as observed by the current user there during time slot $t$. More specifically, $y_i(t)=1$ tells that the current user at time $t$ observes a hazard (e.g., ‘black ice’ segments, poor visibility, jamming) after choosing path $\pi(t)=i$. $y_i(t)=0$ tells that the user does not observe any hazard on path $i$. Finally, $y_i(t)=\emptyset$ tells that this user travels on another path with $\pi(t)\neq i$, without making any observation of path $i$. 

Given $\pi(t)=i$, the chance for the user to observe $y_i(t)=1$ or $0$ depends on the random correlation coefficient $\alpha_i(t)$. Under the correlation state $\alpha_i(t)=\alpha_H$ or $\alpha_L$ at time $t$, we respectively denote the probabilities for the user to observe a hazard as:
\begin{equation}
    \begin{aligned}
        p_H&=\text{Pr}\big(y_i(t)=1|\alpha_i(t)=\alpha_H\big), \\
        p_L&=\text{Pr}\big(y_i(t)=1|\alpha_i(t)=\alpha_L\big).
    \end{aligned}\label{p_H}
\end{equation}
Note that $p_L<p_H$ because a risky path in bad traffic condition ($\alpha_i(t)=\alpha_H$) has a larger probability for the user to observe a hazard (i.e., $y_i(t)=1$). Even if path $i$ has good traffic condition ($\alpha_i(t)=\alpha_L$), it is not entirely hazard free and there is still some probability $p_L$ to face a hazard.

As users keep learning and sharing traffic conditions with the crowdsourcing platform, the historical data of their observations $(\mathbf{y}(1),\cdots,\mathbf{y}(t-1))$ and routing decisions $(\pi(1),\cdots,\pi(t-1))$ before time $t$ keep growing in the time horizon. To simplify the ever-growing history set, we equivalently translate these historical observations into a hazard belief $x_i(t)$ for seeing bad traffic condition $\alpha_i(t)=\alpha_H$ at time $t$, by using the Bayesian inference:
\begin{equation}
    x_i(t)=\text{Pr}\big(\alpha_i(t)=\alpha_H|x_i(t-1),\pi(t-1),\mathbf{y}(t-1)\big).\label{def_x}
\end{equation}
Given the prior probability $x_i(t)$, the platform will further update it to a posterior probability $x_i'(t)$ after a new user with routing decision $\pi(t)$ shares his observation $y_i(t)$ during the time slot:
\begin{equation}
    x_i'(t)=\text{Pr}\big(\alpha_i(t)=\alpha_H|x_i(t),\pi(t),\mathbf{y}(t)\big).\label{def_x'}
\end{equation}
Below, we explain the dynamics of our information learning model.
\begin{itemize}
    \item At the beginning of time slot $t$, the platform publishes any risky path $i$’s hazard belief $x_i(t)$ in (\ref{def_x}) about coefficient $\alpha_i(t)$ and the latest expected latency $\mathbb{E}[\ell_i(t)|x_i(t-1),y_i(t-1)]$ to summarize observation history $(\mathbf{y}(1),\cdots,\mathbf{y}(t-1))$ till $t-1$.

    \item During time slot $t$, a user arrives to choose a path (e.g., $\pi(t)=i$) to travel and reports his following observation $y_i(t)$. Then the platform updates the posterior probability $x_i'(t)$, conditioned on the new observation $y_i(t)$ and the prior probability $x_i(t)$ in (\ref{def_x'}). For example, if $y_i(t)=0$, by Bayes’ Theorem, $x_i'(t)$ for the correlation coefficient $\alpha_i(t)=\alpha_H$ is
    \begin{align}
        x_i'(t)
        =&\text{Pr}\big(\alpha_i(t)=\alpha_H|x_i(t),\pi(t)=i,y_i(t)=0\big)\label{x_i'(t)_y=0}\\
        =&\frac{x_i(t)(1-p_H)}{x_i(t)(1-p_H)+(1-x_i(t))(1-p_L)}.\notag
    \end{align}
    Similarly, if $y(t)=1$, we have
    \begin{equation}
        x_i'(t)=\frac{x_i(t)p_H}{x_i(t)p_H+(1-x_i(t))p_L}.\label{x_i'(t)_y=1}
    \end{equation}
    Besides this traveled path $i$, for any other path $j\in\{1,\cdots,N\}$ with $y_j(t)=\emptyset$, we keep $x'_j(t)=x_j(t)$ as there is no added observation to this path at $t$.

    \item At the end of this time slot, the platform estimates the posterior correlation coefficient:
    \begin{equation}
        \begin{aligned}
            \mathbb{E}[\alpha_i(t)|x'_i(t)]&=\mathbb{E}[\alpha_i(t)|x_i(t),y_i(t)]
            \\&=x'_i(t)\alpha_H+(1-x'_i(t))\alpha_L.
        \end{aligned}\label{E_alpha}
    \end{equation} 
    By combining (\ref{E_alpha}) with (\ref{L_i(t+1)}), we can obtain the expected travel latency on stochastic path $i$ for time $t+1$ as
    \begin{align}
        &\Scale[0.95]{\mathbb{E}[\ell_i(t+1)|x_i(t),y_i(t)]}\label{E[L_i(t+1)]}=\\
        \!\!&\begin{cases}
            \Scale[0.95]{\mathbb{E}[\alpha_i(t)|x'_i(t)] \mathbb{E}[\ell_i(t)|x_i(t-1),y_i(t-1)]+\Delta \ell},\\
            \quad\quad\quad\quad\quad\quad\quad\quad\quad\quad\quad\quad\quad\quad\quad\quad\ \ \ \text{if }\Scale[0.95]{\pi(t)=i},\\
            \Scale[0.95]{\mathbb{E}[\alpha_i(t)|x'_i(t)] \mathbb{E}[\ell_i(t)|x_i(t-1),y_i(t-1)]},\text{ if }\Scale[0.95]{\pi(t)\neq i}.
        \end{cases}\notag
    \end{align}
    Based on the partially observable Markov chain in Fig. \ref{fig:POMDP}, the platform updates each path $i$'s hazard belief from $x_i'(t)$ to $x_i(t+1)$ below: 
    \begin{equation}
        x_i(t+1)=x_i'(t)q_{HH}+\big(1-x_i'(t)\big)q_{LH}.\label{x_i(t)}
    \end{equation}
    Finally, the new time slot $t+1$ begins and repeats the process since above.
\end{itemize}

\section{POMDP Problem Formulations for Myopic and Socially Optimal Policies}\label{section3}
Based on the dynamic congestion and crowdsourcing models in the last section, we formulate the problems of myopic policy (for guiding myopic users' selfish routing) and the socially optimal policy (for the social planner/platform's best path advisory), respectively. 

\subsection{Problem Formulation for Myopic Policy}
In this subsection, we consider the myopic policy (e.g. used by Waze and Google Maps) that the selfish users will naturally follow. First, we summarize the dynamics of expected travel latencies among all $N+1$ paths and the hazard beliefs of $N$ stochastic paths into vectors:
\begin{align}
    &\begin{aligned}
        \mathbf{L}(t)=\big\{&\ell_0(t),\mathbb{E}[\ell_1(t)|x_i(t-1),y_i(t-1)],\cdots,\\&\mathbb{E}[\ell_N(t)|x_N(t-1),y_N(t-1)]\big\},
    \end{aligned}\notag\\
    &\mathbf{x}(t)=\{x_1(t),\cdots,x_N(t)\},\label{LX_set}
\end{align}
which are obtained based on (\ref{E[L_i(t+1)]}) and (\ref{x_i(t)}). For a user arrival at time $t$, the platform provides him with $\mathbf{L}(t)$ and $\mathbf{x}(t)$ to help make his routing decision. We define the best stochastic path $\hat{\iota}(t)$ to be the one out of $N$ risky paths to provide the shortest expected travel latency at time $t$ below:
\begin{equation}
    \hat{\iota}(t)=\arg\min_{i\in\{1,\cdots,N\}} \mathbb{E}[\ell_i(t)|x_i(t-1),y_i(t-1)]. \label{hat_i}
\end{equation}
The selfish user will only choose between safe path 0 and this path $\hat{\iota}(t)$ to minimize his own travel latency. 

We formulate this problem as a POMDP, where the time correlation state $\alpha_i(t)$ of each stochastic path $i$ is partially observable to users in Fig. \ref{fig:POMDP}. Thus, the states here are $\mathbf{L}(t)$ and $\mathbf{x}(t)$ in (\ref{LX_set}). Under the myopic policy, define ${C^{(m)}\big(\mathbf{L}(t), \mathbf{x}(t)\big)}$ to be the long-term discounted cost function with discount factor $\rho<1$ to include social cost of all users since $t$. Then its dynamics per user arrival has the following two cases.
If $\mathbb{E}[\ell_{\hat{\iota}(t)}(t)|x_{\hat{\iota}(t)}(t-1),y_{\hat{\iota}(t)}(t-1)]\geq \ell_0(t)$, a selfish user will choose path 0 and add $\Delta \ell$ to path 0 to have latency $\ell_0(t+1)=\alpha \ell_0(t)+\Delta \ell$ in (\ref{L_0(t+1)}). Since no user enters stochastic path $i$, there is no information reporting (i.e., $y_i(t)=\emptyset$) and $x_i'(t)$ in (\ref{def_x'}) equals $x_i(t)$ in (\ref{def_x}) for updating $x_i(t+1)$ in (\ref{x_i(t)}). The expected travel latency of stochastic path $i$ in the next time slot is updated to $\mathbb{E}[\ell_i(t+1)|x_i(t),y_i(t)=\emptyset]$ according to (\ref{E[L_i(t+1)]}). In consequence, the travel latency and hazard belief sets at the next time slot $t+1$ are updated to
\begin{align}
    &\begin{aligned}\mathbf{L}(t+1)=\big\{&\alpha\ell_0(t)+\Delta\ell, \mathbb{E}[\ell_1(t+1)|x_1(t),y_1(t)=\emptyset],\\&\cdots,\mathbb{E}[\ell_N(t+1)|x_N(t),y_N(t)=\emptyset]\big\},\end{aligned}\notag\\
    &\mathbf{x}(t+1)=\big\{x_1(t+1),\cdots,x_N(t+1)\big\}.\label{L0X0}
\end{align}
Then the cost-to-go $Q_0^{(m)}(t+1)$ since the next user is
\begin{equation}
    \Scale[0.95]{Q_0^{(m)}(t+1)=C^{(m)}\Big(\mathbf{L}(t+1),\mathbf{x}(t+1)\big|y_{\hat{\iota}(t)}(t)=\emptyset\Big)}.\label{Q_0m}
\end{equation}

If $\Scale[0.95]{\mathbb{E}[\ell_{\hat{\iota}(t)}(t)|x_{\hat{\iota}(t)}(t-1),y_{\hat{\iota}(t)}(t-1)]< \ell_0(t)}$, the user will choose the best stochastic path $\hat{\iota}(t)$ in (\ref{hat_i}). Then the platform updates the expected travel latency on path $\hat{\iota}(t)$ to $\Scale[0.95]{\mathbb{E}[\ell_{\hat{\iota}(t)}(t)|x_{\hat{\iota}(t)}(t),y_{\hat{\iota}(t)}(t)]}$ in (\ref{E[L_i(t+1)]}), depending on whether $y_{\hat{\iota}(t)}(t)=1$ or $0$. Note that according to (\ref{p_H}),
\begin{equation}
    \Scale[0.95]{\mathbf{Pr}\big(y_{\hat{\iota}(t)}(t)=1\big)=\big(1-x_{\hat{\iota}(t)}(t)\big)p_L+x_{\hat{\iota}(t)}(t)p_H}. \label{PG}
\end{equation}
While path 0's latency in next time changes to $\alpha\ell_0(t)$, and path $i\neq\hat{\iota}(t)$ has no exploration and its expected latency at time $t+1$ becomes $\mathbb{E}[\ell_i(t+1)|x_i(t),y_i(t)=\emptyset]$. Then the expected cost-to-go since the next user in this case is
\begin{align}
    &Q_{\hat{\iota}(t)}^{(m)}(t+1)=\label{E[C]}\\&\Scale[0.95]{\mathbf{Pr}\big(y_{\hat{\iota}(t)}(t)=1\big)C^{(m)}\Big(\mathbf{L}(t+1),\mathbf{x}(t+1)\big|y_{\hat{\iota}(t)}(t)=1\Big)}\notag\\&+ \Scale[0.95]{\mathbf{Pr}\big(y_{\hat{\iota}(t)}(t)=0\big)C^{(m)}\big(\mathbf{L}(t+1),\mathbf{x}(t+1)\big|y_{\hat{\iota}(t)}(t)=0\Big)}.\notag
\end{align}

To combine (\ref{Q_0m}) and (\ref{E[C]}), we formulate the $\rho$-discounted long-term cost function since time $t$ under myopic policy as
\begin{equation*}
    \Scale[0.95]{C^{(m)}\big(\mathbf{L}(t), \mathbf{x}(t)\big)=}\quad\quad\quad\quad\quad\quad\quad\quad\quad\quad\quad\quad\quad\quad\quad
\end{equation*}
    \begin{numcases}{}
     \Scale[0.97]{\ell_0(t)+\rho Q_{0}^{(m)}(t+1)},\notag\\\quad\quad\ \ \text{if }\Scale[0.97]{\mathbb{E}[\ell_{\hat{\iota}(t)}(t)|x_{\hat{\iota}(t)}(t-1),y_{\hat{\iota}(t)}(t-1)]\geq \ell_0(t)},\notag\\
     \Scale[0.97]{\mathbb{E}[\ell_{\hat{\iota}(t)}(t)|x_{\hat{\iota}(t)}(t-1),y_{\hat{\iota}(t)}(t-1)]+\rho Q_{\hat{\iota}(t)}^{(m)}(t+1)},\notag\\\quad\quad\ \ \text{otherwise.}\label{cost_Cm}
    \end{numcases}
A selfish user is not willing to explore any stochastic path $i$ with longer expected travel latency, and the next arrival may not know the fresh congestion information. On the other hand, selfish users may keep choosing the path with the shortest latency and jamming this path for future users. 

\subsection{Socially Optimal Policy Problem Formulation}
Different from the myopic policy that focuses on the one-shot to minimize the current user's immediate travel cost, the goal of the social optimum is to find optimal policy $\pi^*(t)$ at any time $t$ to minimize the expected social cost over an infinite time horizon.

Denote the long-term $\rho$-discounted cost function by ${C^*\big(\mathbf{L}(t), \mathbf{x}(t)\big)}$ under the socially optimal policy. The optimal policy depends on which path choice yields the minimal long-term social cost. If the platform asks the current user to choose path 0, this user will bear cost $\ell_0(t)$ to travel this path. Due to no information observation (i.e., $\mathbf{y}(t)=\emptyset$), the cost-to-go $Q^*_0(t+1)$ from the next user can be similarly determined as (\ref{Q_0m}) with $\mathbf{L}(t+1)$ and $\mathbf{x}(t+1)$ in (\ref{L0X0}).

If the platform asks the user to explore a stochastic path $i$, this choice is not necessarily path $\hat{\iota}(t)$ in (\ref{hat_i}). Then the platform updates $\mathbf{x}(t+1)$, depending on whether the user's observation on this path is $y_i(t)=1$ or $y_i(t)=0$. Similar to (\ref{E[C]}), the optimal expected cost function from next user is denoted as $Q^*_{i}(t+1)$. Then we are ready to formulate the social cost function under socially optimal policy below:
\begin{align}
    &C^*\big(\mathbf{L}(t), \mathbf{x}(t)\big)\label{cost_C*}\\=&\min_{i\in\{1,\cdots,N\}}\big\{\ell_0(t)+\rho Q^*_0(t+1), \ell_{i}(t)+\rho Q^*_{i}(t+1)\big\}.\notag
\end{align}
Problem (\ref{cost_C*}) is non-convex and its analysis will cause the curse of dimensionality in the infinite time horizon \cite{bellman1966dynamic}. Though it is difficult to solve, we still analytically compare the two policies by their structural results below.

\section{Comparing Myopic Policy to Social Optimum for PoA Analysis}\label{section4}
In this section, we first prove that both myopic and socially optimal policies to explore stochastic paths are of threshold-type with respect to expected travel latency. Then we show that the myopic policy may both under-explore and over-explore risky paths. \footnote{Over/under exploration means that myopic policy will choose risky path $i$ more/less often than what the social optimum suggests.} Finally, we prove that the myopic policy can perform arbitrarily bad.

\begin{lemma}\label{lemma:monotonocity}
The cost functions $C^{(m)}\big(\mathbf{L}(t), \mathbf{x}(t)\big)$ in (\ref{cost_Cm}) and $C^*\big(\mathbf{L}(t), \mathbf{x}(t)\big)$ in (\ref{cost_C*}) under both policies increase with any path's expected latency $\mathbb{E}[\ell_i(t)|x_i(t-1),y_i(t-1)]$ in $\mathbf{L}(t)$ and $\mathbf{x}(t)$ in (\ref{LX_set}).
\end{lemma}
With this monotonicity result, we next prove that both policies are of threshold-type.
\begin{proposition}\label{prop:threshold}
Provided with $\mathbf{L}(t)$ and $\mathbf{x}(t)$ in (\ref{LX_set}), the user arrival at time $t$ under the myopic policy keeps staying with path 0, until the expected latency of the best stochastic path $\hat{\iota}(t)$ in (\ref{hat_i}) reduces to be smaller than the following threshold:
\begin{equation}
    \ell^{(m)}(t)=\ell_0(t). \label{L_mopic}
\end{equation}
Similarly, the socially optimal policy will choose stochastic path $i$ instead of path 0 if $\mathbb{E}[\ell_i(t)|x_i(t-1),y_i(t-1)]$ is less than the following threshold:
\begin{equation}
    \begin{aligned}
    \Scale[0.95]{\ell_i^*(t)=\arg \max_{z}\big\{z|z \leq \rho Q^*_{i}(t+1)-\rho Q^*_0(t+1)-\ell_0(t)\big\},}
    \end{aligned}\label{L_optimal}
\end{equation}
which increases with hazard belief $x_{i}(t)$ of risky path $i$.
\end{proposition}
Let $\pi^{(m)}(t)$ and $\pi^*(t)$ denote the routing decisions at time $t$ under myopic and socially optimal policies, respectively. We next compare the exploration thresholds $\ell^{(m)}(t)$ and $\ell_i^{*}(t)$ as well as their associated social costs. 
\begin{lemma}\label{lemma:explore}
If $\pi^{(m)}(t)\neq \pi^*(t)$, then the expected travel latencies on these two chosen paths by the two policies satisfy
\begin{align}
    &\mathbb{E}[\ell_{\pi^*(t)}(t)|\mathbf{x}(t-1),\mathbf{y}(t-1)]\notag\\ \leq& \frac{1}{1-\rho}\mathbb{E}[\ell_{\pi^{(m)}(t)}(t)|\mathbf{x}(t-1),\mathbf{y}(t-1)].\label{lemma_inequality}
\end{align}
\end{lemma}
Intuitively, if the current travel latencies on different paths obviously differ, the two policies tend to make the same routing decision. (\ref{lemma_inequality}) is more likely to hold for large $\rho$.

Next, we define the stationary belief $x_i(t)$ of high hazard state $\alpha_H$ as $\Bar{x}$, and we provide it below by using steady-state analysis of Fig. \ref{fig:POMDP}:
\begin{equation}
    \Bar{x}=\frac{1-q_{LL}}{2-q_{LL}-q_{HH}}.\label{x_bar}
\end{equation}
Based on Proposition \ref{prop:threshold} and Lemma \ref{lemma:explore}, we analytically compare the two policies below.
\begin{proposition}\label{prop:explore}
There exists a belief threshold $x^{th}$ satisfying
\begin{equation}
    \min\Big\{\frac{\alpha-\alpha_L}{\alpha_H-\alpha_L},\Bar{x}\Big\}\leq x^{th}\leq \max\Big\{\frac{\alpha-\alpha_L}{\alpha_H-\alpha_L},\Bar{x}\Big\}.\label{x_th}
\end{equation}
As compared to socially optimal policy, if risky path $i\in\{1,\cdots,N\}$ has weak hazard belief $x_i(t)< x^{th}$, myopic users will only over-explore this path with $\ell^{(m)}(t)\geq \ell_i^*(t)$. If strong hazard belief with $x_i(t)>x^{th}$, myopic users will only under-explore this path with $\ell^{(m)}(t)\leq \ell_i^*(t)$.
\end{proposition} 
Here $\frac{\alpha-\alpha_L}{\alpha_H-\alpha_L}$ in (\ref{x_th}) is derived by equating path $i$'s expected coefficient $\mathbb{E}[\alpha_i(t)|x'_i(t)]$ in (\ref{E_alpha}) to path 0's $\alpha$. Proposition \ref{prop:explore} tells that the myopic policy misses both exploitation and exploration over time. If the hazard belief on path $i\in\{1,\cdots,N\}$ is weak (i.e., $x_i(t)<x^{th}$), myopic users choose stochastic path $i$ without considering the congestion to future others on the same path. While the the socially optimal policy may still recommend users to safe path 0 to further reduce the congestion cost on path $i$ for the following user. On the other hand, if $x_i(t)>x^{th}$, the socially optimal policy may still want to explore path $i$ to exploit hazard-free state $\alpha_L$ on this path for future use. This result is also consistent with $\ell_i^*(t)$'s monotonicity in $x_i(t)$ in Proposition \ref{prop:threshold}. 

\begin{figure}[t]
    \centering
    \includegraphics[width=0.28\textwidth]{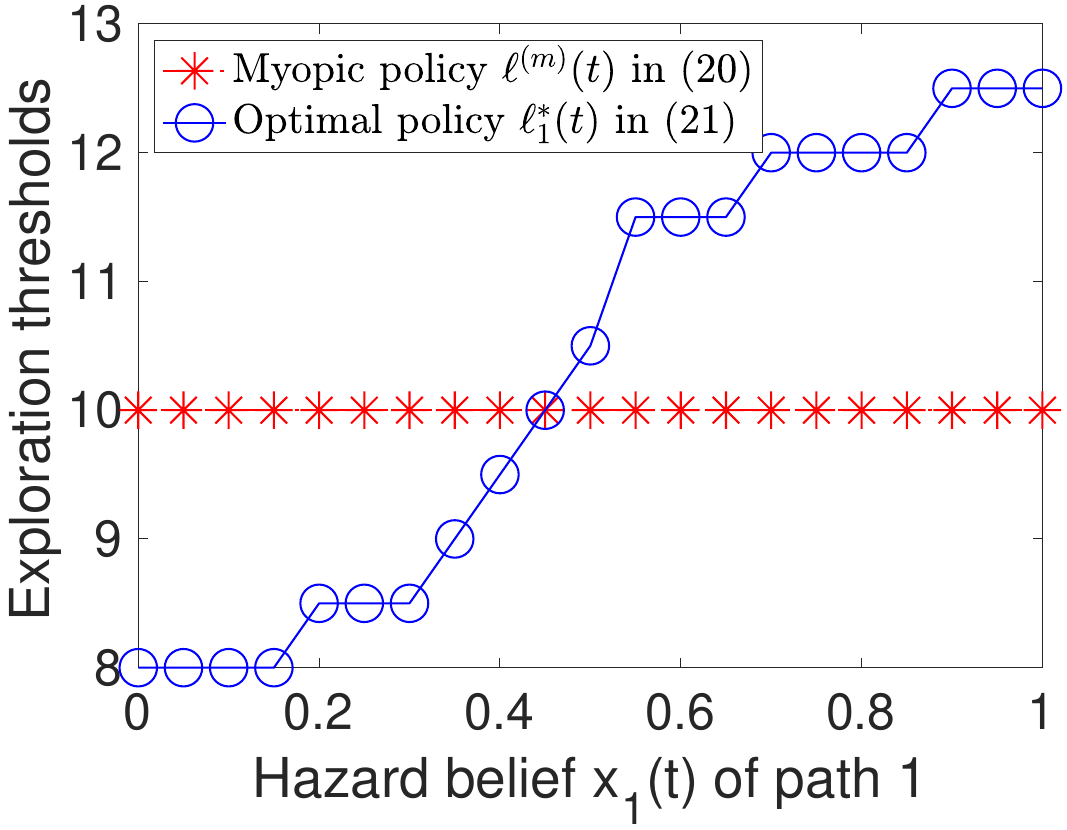}
    \caption{The socially optimal policy's exploration threshold $\ell_1^*(t)$ and myopic policy's threshold $\ell^{(m)}(t)$ versus hazard belief $x_1(t)$ in a two-path transportation network with $N=1$.  We set $\alpha=0.6,\alpha_H=1.2,\alpha_L=0.2,q_{LL}=0.5,q_{HH}=0.5,\Delta\ell=2,p_H=0.8,p_L=0.3,\ell_0(t)=10$ and $x_1(t)=0.1$ at current time $t$.}
    \label{fig:exploration_x}
\end{figure}

In Fig. \ref{fig:exploration_x}, we simulate Fig. \ref{fig:congestion_game} using a simple two-path transportation network with $N=1$. We plot exploration thresholds $\ell^{(m)}(t)$ in (\ref{L_mopic}) under myopic policy and optimal $\ell_1^*(t)$ in (\ref{L_optimal}) versus hazard belief $x_1(t)$ of path 1. These two thresholds are very different in Fig. \ref{fig:exploration_x}. Given the belief threshold $x^{th}=0.45$ here, if the hazard belief $x_1(t)<x^{th}$, we have the myopic exploration threshold $\ell^{(m)}(t)>\ell_1^*(t)$ to over-explore stochastic path. If $x_i(t)>x^{th}$, the myopic exploration threshold satisfies $\ell^{(m)}(t)<\ell_1^*(t)$ to over-explore. This result is consistent with Proposition \ref{prop:explore}.

After comparing the two policies' thresholds, we are ready to further examine their performance gap. Following \citeauthor{koutsoupias1999worst} (\citeyear{koutsoupias1999worst}), we define the price of anarchy (PoA) to be the maximum ratio between the social cost under myopic policy in (\ref{cost_Cm}) and the minimal social cost in (\ref{cost_C*}), by searching all possible system parameters:
\begin{align}
    &\text{PoA}^{(m)}=\max_{\substack{\alpha,\alpha_H,\alpha_L,q_{LL},q_{HH},\\\mathbf{x}(t),\mathbf{L}(t),\Delta \ell, p_H, p_L}}{\frac{C^{(m)}\big(\mathbf{L}(t), \mathbf{x}(t)\big)}{C^{*}\big(\mathbf{L}(t), \mathbf{x}(t)\big)}},\label{PoA}
\end{align}
which is obviously larger than 1. Then we present the lower bound of PoA in the following proposition.
\begin{proposition}\label{thm:poa}
As compared to the social optimum in (\ref{cost_C*}), the myopic policy in (\ref{cost_Cm}) achieves $\text{PoA}^{(m)}\geq \frac{1}{1-\rho}$, which can be arbitrarily large for discount factor $\rho\rightarrow 1$.
\end{proposition}
In this worst-case PoA analysis, we consider a two-path network example, where the myopic policy always chooses safe path 0 but the socially optimal policy frequently explores stochastic path 1 to learn $\alpha_L$. Here we initially set $\ell_0(0)=\frac{\Delta \ell}{1-\alpha}$ such that the travel latency $\ell_0(1)=\alpha\ell_0(0)+\Delta \ell$ in (\ref{L_0(t+1)}) equals $\frac{\Delta\ell}{1-\alpha}$ all the time for myopic users. Without myopic users' routing on stochastic path 1, we also keep the expected travel latency on stochastic path 1 unchanged, by setting $x_1(0)=\Bar{x}$ in (\ref{x_bar}) and $\mathbb{E}[\alpha_1(0)|x_1(0)=\Bar{x}]=1$ in (\ref{E_alpha}). Then a myopic user at any time $t$ will never explore the stochastic path 1 given $\ell_1(t)=\ell_0(t)$, resulting in the social cost to be $\frac{\ell_0(0)}{1-\rho}$ in the infinite time horizon. However, the socially optimal policy frequently asks a user arrival to explore path 1 to learn a good condition ($\alpha_L=0$) for following users. We make $q_{LL}\rightarrow 1$ to maximally reduce the travel latency of path 1, and the optimal social cost is thus no more than $\ell_1(0)+\frac{\rho}{1-\rho}\Delta\ell$. Letting $\frac{\Delta\ell}{\ell_0(0)}\rightarrow 0$, we obtain $\text{PoA}^{(m)}\geq \frac{1}{1-\rho}$.

By Proposition \ref{thm:poa}, the myopic policy performs worse, as discount factor $\rho$ increases and future costs become more important. As $\rho\rightarrow 1$, PoA approaches infinity and the learning efficiency in the crowdsourcing platform
becomes arbitrarily bad to opportunistically reduce the congestion. Thus, it is critical to design efficient incentive mechanism to greatly reduce the social cost.

\section{Selective Information Disclosure}\label{section5}
To motivate a selfish user to follow the optimal path advisory at any time, we need to design a non-monetary information mechanism, which naturally satisfies budget balance and is easy to implement without enforcing monetary payments. Our key idea is to selectively disclose the latest expected travel latency set $\mathbf{L}(t)$ of all paths, depending on a myopic user's intention to over- or under-explore stochastic paths at time $t$. To avoid users from perfectly inferring $\mathbf{L}(t)$, we purposely hide the latest hazard belief set $\mathbf{x}(t)$, routing history $\big(\pi(1),\cdots,\pi(t-1)\big)$, and past traffic observation set $\big(\mathbf{y}(1),\cdots, \mathbf{y}(t-1)\big)$, but always provide socially optimal path recommendation $\pi^*(t)$ to any user.
Provided with selective information disclosure, we allow sophisticated users to reverse-engineer the path latency distribution and make selfish routing under our mechanism. 

Before formally introducing our selective information disclosure in Definition \ref{def:mechanism}, we first consider an information hiding policy $\pi^{\emptyset}(t)$ as a benchmark. Similar information hiding mechanisms were proposed and studied in the literature (e.g., \citeauthor{tavafoghi2017informational} \citeyear{tavafoghi2017informational} and \citeauthor{li2019recommending} \citeyear{li2019recommending}). In this benchmark mechanism, the user without any information believes that the expected hazard belief $x_i(t)$ of any stochastic path $i\in \{1,\cdots,N\}$ has converged to its stationary hazard belief $\Bar{x}$ in (\ref{x_bar}). Then he can only decide his routing policy $\pi^{\emptyset}(t)$ by comparing $\alpha$ of safe path 0 to $\mathbb{E}[\alpha_i(t)|\Bar{x}]$ in (\ref{E_alpha}) of any path $i$. 
\begin{proposition}\label{lemma:pi^empty}
Given no information from the platform, a user arrival at time $t$ uses the following routing policy: 
\begin{equation}
    \pi^{\emptyset}(t)=\begin{cases}
    0, &\text{if }\Bar{x}\geq \frac{\alpha-\alpha_L}{\alpha_H-\alpha_L},\\
    i\text{ w/ probability }\frac{1}{N},&\text{if }\Bar{x}< \frac{\alpha-\alpha_L}{\alpha_H-\alpha_L},
    \end{cases}\label{pi^empty}
\end{equation}
where $i\in\{1,\cdots,N\}$. This hiding policy leads to $\text{PoA}^{\emptyset}=\infty$, regardless of discount factor $\rho$.
\end{proposition}
Even if we still recommend optimal routing $\pi^*(t)$ in (\ref{cost_C*}), a selfish user sticks to some risky path $i$ given low hazard belief $\Bar{x}<\frac{\alpha-\alpha_L}{\alpha_H-\alpha_L}$. This hiding policy can differ a lot from the socially optimal policy in (\ref{cost_C*}) since users cannot observe the latest travel latencies. To tell the $\text{PoA}^{\emptyset}=\infty$, we consider the simplest two-path network example: initially safe path 0 has $\ell_0(t=0)=0$ with $\alpha\rightarrow 1$, and risky path 1 has an arbitrarily large travel latency $\ell_1(0)$ with $\Bar{x}=0$ and $\mathbb{E}[\alpha_1(t)|\Bar{x}]= 0$, by letting $q_{LL}=1$ and $\alpha_L=0$. Given $\mathbb{E}[\alpha_1(t)|\Bar{x}]<\alpha$ or simply $\Bar{x}<\frac{\alpha-\alpha_L}{\alpha_H-\alpha_L}$, a selfish user always chooses path $\pi^{\emptyset}(t)=1$, leading to social cost $\ell_1(0)+\frac{\rho\Delta\ell}{1-\rho}$. While letting the first user exploit $\ell_0(0)=0$ of path 0 to reduce $\mathbb{E}[\ell_1(1)|\Bar{x},\emptyset]$ to 0 for path 1 at time $1$, the socially optimal cost is thus $\frac{\rho^2\Delta\ell}{1-\rho}$. Letting $\frac{(1-\rho)\ell_1(0)}{\rho^2\Delta\ell}\rightarrow\infty$, we obtain $\text{PoA}^{\emptyset}=\infty$. 

This is a $\text{PoA}^{\emptyset}$ example with the maximum-exploration of stochastic paths, which is opposite to the zero-exploration $\text{PoA}^{(m)}$ example after Proposition \ref{thm:poa}. 
Given neither information hiding policy $\pi^{\emptyset}(t)$ nor myopic policy $\pi^{(m)}(t)$ under full information sharing works well, we need to design an efficient mechanism to selectively disclose information to users to reduce the social cost. 
\begin{definition}{(\textbf{Selective Information Disclosure (SID) Mechanism:})}\label{def:mechanism}
If a user arrival at time $t$ is expected to choose a different route $\pi^{\emptyset}(t)\neq 0$ in (\ref{pi^empty}) from optimal $\pi^*(t)=0$ in (\ref{cost_C*}), then our SID mechanism will disclose the latest expected travel latency set $\mathbf{L}(t)$ to him. Otherwise, our mechanism hides $\mathbf{L}(t)$ from this user. Besides, our mechanism always provides optimal path recommendation $\pi^*(t)$, without sharing hazard belief set $\mathbf{x}(t)$, routing history $\big(\pi(1),\cdots,\pi(t-1)\big)$, or past observation set $\big(\mathbf{y}(1),\cdots, \mathbf{y}(t-1)\big)$.
\end{definition}
According to Definition \ref{def:mechanism}, if $\pi^*(t)=0$ but a user at time $t$ makes routing decision $\pi^{\emptyset}(t)\neq 0$ under $\Bar{x}<\frac{\alpha-\alpha_L}{\alpha_H-\alpha_L}$ in (\ref{pi^empty}), our mechanism discloses $\mathbf{L}(t)$ to avoid him from choosing any stochastic path with large expected travel latency. In the other cases, we simply hide $\mathbf{L}(t)$ from any user arrival, as the user already follows optimal routing $\pi^*(t)$. 

In consequence, the worst-case for our SID mechanism only happens when $\pi^{\emptyset}(t)\neq 0$ and $\pi^*(t)=0$ under $\Bar{x}<\frac{\alpha-\alpha_L}{\alpha_H-\alpha_L}$ in (\ref{pi^empty}). We still consider the same two-path network example with the maximum-exploration after Proposition \ref{lemma:pi^empty} to show why this SID mechanism works. In this example, our mechanism will provide $\mathbf{L}(t)$, including $\ell_0(0)$ and $\ell_1(0)$, to each user arrival. Observing huge $\ell_1(0)$, the first user turns to choose path $0$ with $\ell_0(0)=0$, which successfully avoids the infinite social cost under $\pi^{\emptyset}(t)$. Furthermore, our SID mechanism successfully avoids the worst-cases of $\text{PoA}^{(m)}$ in Proposition \ref{thm:poa}. Next we prove that our mechanism well bounds the PoA in the following.
\begin{theorem}\label{thm:poa_incentive}
Our SID mechanism results in $\text{PoA}^{(\text{SID})}\leq \frac{1}{1-\frac{\rho}{2}}$, which is always no more than $2$.
\end{theorem}
In the worst-case of $\pi^{\emptyset}(t)\neq 0$ and $\pi^*(t)=0$ for our SID mechanism's $\text{PoA}^{(\text{SID})}$, a user knowing $\mathbf{L}(t)$ may deviate to follow the myopic policy $\pi^{(m)}(t)\neq 0$ in (\ref{cost_Cm}).
To explain the bounded $\text{PoA}^{(\text{SID})}$, we consider a two-path network example with the maximum-exploration under the myopic policy. Here we start with $\ell_0(0)=\ell_1(0)-\varepsilon$ for safe path 0 with $\alpha\rightarrow 1$ to keep the travel latency on path 0 unchanged if no user chooses that path, where $\varepsilon$ is positive infinitesimal. We set $\ell_1(0)=\frac{\Delta \ell}{1-\mathbb{E}[\alpha_1(0)|\Bar{x}]}$ for stochastic path 1 with $x_1(0)=\Bar{x}$, such that the travel latency $\mathbb{E}[\ell_1(t)|\Bar{x},y_1(t-1)]$ equals $\ell_1(0)$ all the time if all users choose that path.
Then in this system, users keep choosing path 1 under myopic policy $\pi^{(m)}(t)$ in (\ref{cost_Cm}) to receive social cost $\frac{\ell_1(0)}{1-\rho}$. However, the socially optimal policy may want the first user to exploit path 0 to permanently reduce path 1's expected travel latency for following users there. Thanks to the first user's routing of path 0, the expected travel latency for each following user choosing path 1 at time $t$ is greatly reduced to be less than $\ell_1(0)$ yet is still no less than $\frac{\ell_1(0)}{2}$ for non-zero $\mathbb{E}[\alpha_1(t)|\Bar{x}]$. Then the minimum social cost is reduced to be no less than $\ell_0(0)+\frac{\rho \ell_1(0)}{2-2\rho}$, leading to $\text{PoA}^{\text{(SID)}}\leq \frac{1}{1-\frac{\rho}{2}}$.

\begin{figure}[t]
    \centering
    \includegraphics[width=0.3\textwidth]{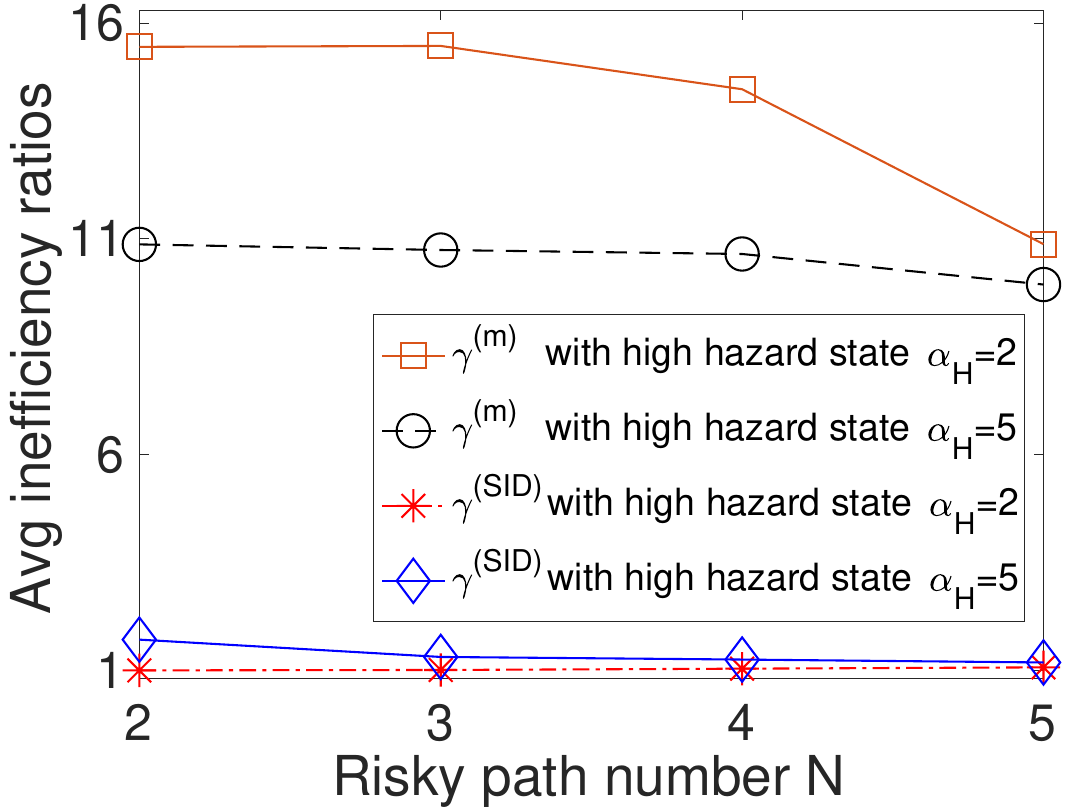}
    \caption{Average inefficiency ratios $\gamma^{(m)}$ under myopic policy in (\ref{cost_Cm}) and $\gamma^{(\text{SID})}$ under our selective information disclosure. We vary risky path number $N$ in set $\{2,3,4,5\}$. We set $\alpha=0.99,\alpha_L=0,\Delta \ell=1, p_H=0.8, p_L=0.2, q_{HH}=0.99,q_{LL}=0.99$ here, and we change $\alpha_H=2$ and $\alpha_H=5$ to make comparison. At initial time $t=0$, we let $\ell_0(0)=100,\ell_i(0)=105$ and $x_i(0)=0.5$ for any path.}
    \label{fig:avg_ratio}
\end{figure}

Besides the worst-case performance analysis, we further verify our mechanism's average performance using extensive simulations. Define the following average inefficiency ratio between expected social costs achieved by our SID mechanism and social optimum in (\ref{cost_C*}):
\begin{equation}
\gamma^{(\text{SID})}=\frac{\mathbb{E}\big[C^{(\text{SID})}\big(\mathbf{L}(t),\mathbf{x}(t)\big)\big]}{\mathbb{E}\big[C^{*}\big(\mathbf{L}(t),\mathbf{x}(t)\big)\big]}.
\end{equation}
To compare, we define $\gamma^{(m)}$ to be the average inefficiency ratio between social costs achieved by the myopic policy in (\ref{cost_Cm}) and socially optimal policy in (\ref{cost_C*}). After running $50$ long-term experiments for averaging each ratio, we plot Fig.~\ref{fig:avg_ratio} to compare $\gamma^{(m)}$ to $\gamma^{(\text{SID})}$ versus risky path number $N$. Fig. \ref{fig:avg_ratio} shows that our SID mechanism obviously reduces $\gamma^{(m)}>10$ to $\gamma^{(\text{SID})}<2$ at $N=2$, which is consistent with Theorem \ref{thm:poa_incentive}. Fig. \ref{fig:avg_ratio} also shows that the efficiency loss due to users' selfish routing decreases with $N$, as more choices of risky paths help negate the hazard risk at each path. Here we also vary high hazard state $\alpha_H$ to make a comparison, and we see that a larger $\alpha_H$ causes less efficiency loss due to users' reduced explorations to risky paths. 

We can also show using simulations that the average inefficiency ratio under information hiding mechanism in Proposition \ref{lemma:pi^empty} has a big gap compared to our SID mechanism, especially when users over-explore with $\Bar{x}<\frac{\alpha-\alpha_L}{\alpha_H-\alpha_L}$. 

\section{Conclusion}\label{section6}
In this paper, we studied how to incentive selfish users to reach the best exploitation-exploration trade-off. We use the POMDP techniques to summarize the congestion probability into a dynamic hazard belief. By considering a simple but fundamental parallel routing network with one deterministic path and multiple stochastic paths for atomic users, we proved that the myopic policy's price of anarchy (PoA) is larger than $\frac{1}{1-\rho}$, which can be arbitrarily large as $\rho\rightarrow1$. To remedy such huge efficiency loss, we proposed a selective information disclosure (SID) mechanism: we only reveal the latest traffic information to users when they intend to over-explore stochastic paths, while hiding such information when they under-explore. We proved that our mechanism reduces PoA to be less than $\frac{1}{1-\frac{\rho}{2}}$. We further examined our mechanism's average-case performance by extensive simulations. We can also extend our system model and key results to a chain road network.

\section{Acknowledgments}
The work was supported by the Ministry of Education, Singapore, under its Academic Research Fund Tier 2 Grant under Award MOE- T2EP20121-0001. 

% \bibliography{aaai22}

\appendix
\section{Proof of Lemma 1}
We only need to prove that $C^{(m)}\big(\mathbf{L}(t), \mathbf{x}(t)\big)$ increases with any path's expected latency $\mathbb{E}[\ell_i(t)|x_i(t-1),y_i(t-1)]$ in $\mathbf{L}(t)$ and $\mathbf{x}(t)$. Then the monotonicity of $C^*\big(\mathbf{L}(t), \mathbf{x}(t)\big)$ similarly holds.

We first prove the monotonicity of $C^{(m)}\big(\mathbf{L}(t), \mathbf{x}(t)\big)$ with respect to a single risky path 1's expected latency $\mathbf{L}(t)=\mathbb{E}[\ell_1(t)|x_i(t-1),y_i(t-1)]$. Let $\mathbf{L}_a(t)=\mathbb{E}[\ell_1(t)|x_i(t-1),y_i(t-1)]$ and $\mathbf{L}_b(t)=\mathbb{E}[\ell_1(t)+1|x_i(t-1),y_i(t-1)]$ denote the two travel latency of risky path 1, respectively. 

According to (\ref{PG}), the probability $P\big(y_{\hat{\iota}(t)}(t)=1\big)$ is always the same for $\mathbf{L}_a(t)$ and $\mathbf{L}_b(t)$. Thus, we have $\mathbf{L}_a(\tau)<\mathbf{L}_b(\tau)$ for any $\tau> t$ based on (\ref{E[L_i(t+1)]}) and $\mathbf{L}_a(t)<\mathbf{L}_b(t)$ at current $t$. While in the safe path, the travel latency are always the same. In consequence, $C^{(m)}\big(\mathbf{L}_a(t), \mathbf{x}(t)\big)\leq C^{(m)}\big(\mathbf{L}_b(t), \mathbf{x}(t)\big)$ is true. It also holds for multiple risky paths.

Next, we prove that $C^{(m)}\big(\mathbf{L}(t), \mathbf{x}(t)\big)$ increases with $\mathbf{x}(t)$. Since a larger $x_i(t)$ causes a higher state $\mathbb{E}[\alpha_i(t)|x'_i(t)]$ in (\ref{E_alpha}), the future expected travel latency in this risky path $i$ $\mathbb{E}[\ell_i(t)+1|x_i(t-1),y_i(t-1)]$ increases with $x_i(t)$ based on (\ref{E[L_i(t+1)]}). Hence, $C^{(m)}\big(\mathbf{L}(t), \mathbf{x}(t)\big)$ will also increase.

This complete the proof. We can use the same method to prove that $C^*\big(\mathbf{L}(t), \mathbf{x}(t)\big)$ holds the same monotonicity as $C^{(m)}\big(\mathbf{L}(t), \mathbf{x}(t)\big)$. Note that one can also prove Lemma 1 using Bellman equation techniques.

\section{Proof of Proposition 1}
It is straightforward to derive the exploration thresholds $\ell^{(m)}(t)$ for myopic policy and $\ell_i^*(t)$ for socially optimal policy, so we only prove that $\ell_i^*(t)$ increases with $x_i(t)$ here.

From the expression of $\ell_i^*(t)$, if we prove
\begin{equation}
    \frac{\partial \Big(Q_0^*(t+1)-Q^*_{i}(t+1)\Big)}{\partial x_i(t)}\geq 0, \label{partial_x}
\end{equation}
we can say that $\ell_i^*(t)$ increases with $x_i(t)$. We first formulate $\frac{\partial Q_0^*(t+1)}{\partial x_{\hat{\iota}_1}(t)}$ in (\ref{partial_Q0}) and $\frac{\partial Q_{\hat{\iota}_1}^*(t+1)}{\partial x_{\hat{\iota}_1}(t)}$ in (\ref{partial_Qi}), respectively. Then we will apply mathematical induction to prove (\ref{partial_x}) based on the two formulations.

Denote the current optimal path by $\hat{\iota}_1$ for latter use. Since the dynamics of travel latency on any path $i\neq \hat{\iota}_1$ is not related with $x_{\hat{\iota}_1}(t)$, $\frac{\partial \ell_i(t)}{\partial x_{\hat{\iota}_1}(t)}$ always equals 0. In consequence, according to the definitions of $Q_0^*(t+1)$ and $Q^*_{i}(t)$, we have
\begin{align}
    \frac{\partial Q_0^*(t+1)}{\partial x_{\hat{\iota}_1}(t)}=&\rho^{T_0}\frac{\partial \mathbb{E}_{T_0}\big[\ell_{\hat{\iota}_1}(t+T_0)|\pi^*(t)=0\big]}{\partial x_{\hat{\iota}_1}(t)}\notag\\&+\rho^{T_0+1} \frac{\partial Q_{\hat{\iota}_1}(t+T_0+1)}{\partial x_{\hat{\iota}_1}(t)},\label{partial_Q0}
\end{align}
where $T_0$ is the elapsed time slots until the next exploration to path $\hat{\iota}_1$ for $Q_{0}^*(t+1)$ since time $t$, and $\mathbb{E}[\ell_{\hat{\iota}_1}(t+T_0)|\pi^*(t)=0]$ is the expected travel latency at $t+T_0$ conditional on $\pi^*(t)=0$. Similarly, we can obtain 
\begin{align}
    \frac{\partial Q_{\hat{\iota}_1}^*(t+1)}{\partial x_{\hat{\iota}_1}(t)}=&\rho^{T_1}\frac{\partial \mathbb{E}_{T_1}\big[\ell_{\hat{\iota}_1}(t+T_1)|\pi^*(t)=\hat{\iota}_1\big]}{\partial x_{\hat{\iota}_1}(t)}\notag\\&+\rho^{T_1+1} \frac{\partial Q_{\hat{\iota}_1}(t+T_1+1)}{\partial x_{\hat{\iota}_1}(t)},\label{partial_Qi}
\end{align}
where $T_1$ is the elapsed time slots until the next exploration to path $\hat{\iota}_1$ for $Q_{\hat{\iota}_1}^*(t)$ after time $t$, and $\mathbb{E}[\ell_{\hat{\iota}_1}(t+T_1)|\pi^*(t)=\hat{\iota}_1]$ is the travel latency at $t+T_1$ conditional on $\pi^*(t)=\hat{\iota}_1$. Note that $T_0\leq T_1$ because the exploration to path $\hat{\iota}_1$ at time $t$ increases the travel latency by $\Delta \ell$, making latter users less willing to explore it. Based on formulations (\ref{partial_Q0}) and (\ref{partial_Qi}), we next use mathematical induction to prove (\ref{partial_x}).

If the time horizon $T=1$, (\ref{partial_x}) is obviously true because $T_0\leq T_1$. Note that if $T_0=T_1=1$, then $ \frac{\partial Q_0^*(t+1)}{\partial x_{\hat{\iota}_1}(t)}- \frac{\partial Q_{\hat{\iota}_1}^*(t+1)}{\partial x_{\hat{\iota}_1}(t)}=0$. We suppose (\ref{partial_x}) is still true for a larger time horizon $T$, where $T\gg T_1\geq T_0$. We next verify that (\ref{partial_x}) is still true for time horizon $T+1$. Since $T\gg T_1\geq T_0$, we only need to compare $\rho^{T_0+1} \frac{\partial Q_{\hat{\iota}_1}(t+T_0+1)}{\partial x_{\hat{\iota}_1}(t)}$ in (\ref{partial_Q0}) to $\rho^{T_1+1} \frac{\partial Q_{\hat{\iota}_1}(t+T_1+1)}{\partial x_{\hat{\iota}_1}(t)}$ in (\ref{partial_Qi}). Since the left time slots for cost-to-go $Q_{\hat{\iota}_1}(t+T_0+1)$ and $Q_{\hat{\iota}_1}(t+T_1+1)$ are $T-T_0$ and $T-T_1$, respectively, we can obtain that 
\[\rho^{T_0+1} \frac{\partial Q_{\hat{\iota}_1}(t+T_0+1)}{\partial x_{\hat{\iota}_1}(t)}-\rho^{T_1+1} \frac{\partial Q_{\hat{\iota}_1}(t+T_1+1)}{\partial x_{\hat{\iota}_1}(t)}\geq 0\]
in (\ref{partial_Q0}) and (\ref{partial_Qi}) based on the conclusion that (\ref{partial_x}) is true when its time horizon is $T-T_0$.

In summary, we show that (\ref{partial_x}) holds for any time horizon $T$. Then we finish the proof that $\ell_i^*(t)$ increases with $x_i(t)$.

\section{Proof of Lemma 2}
As the stochastic path number $N$ increases, $\ell^*_i(t)$ and $\ell^{(m)}(t)$ will approach to identical because more paths help negate the congestion. Hence, we consider the worst-case with two-path transportation network. Under $N=1$, if we prove that (\ref{lemma_inequality}) is always true in the multiple path network with $N\geq 2$. 

Take $\pi^*(t)=0$ and $\pi^{(m)}(t)=1$ as an example, where $\ell_0(t)$ is no less than $\mathbb{E}[\ell_1(t)|x_1(t-1),y_1(t-1)]$. Define $\Delta C_1$ to be the extra travel latency of choosing path 0 instead of path 1 for the current user, which is
\begin{equation*}
    \Delta C_1=\ell_0(t)-\mathbb{E}[\ell_1(t)|x_1(t-1),y_1(t-1)].
\end{equation*}
We also define $\Delta C_2$ to be the exploitation benefit for latter users, which has an upper bound
\begin{align*}
    \Delta C_2 \leq& \rho \mathbb{E}[\ell_1(t)|x_1(t-1),y_1(t-1)]\\&+\rho^2 \mathbb{E}[\ell_1(t)|x_1(t-1),y_1(t-1)]+\cdots\\
    =& \frac{\rho\mathbb{E}[\ell_1(t)|x_1(t-1),y_1(t-1)]}{1-\rho},
\end{align*}
because the travel latency for each user after time $t$ can be reduced at most $\mathbb{E}[\ell_1(t)|x_1(t-1),y_1(t-1)]$. To make sure that $\pi^*(t)=0$ leads to the minimal social cost, we need
\begin{equation*}
    \Delta C_1 \leq \Delta C_2,
\end{equation*}
such that the latter benefit can negate the current extra travel cost. By solving the above equality, we finally obtain that
\begin{equation*}
    \ell_0(t)\leq \frac{1}{1-\rho} \mathbb{E}[\ell_1(t)|x_1(t-1),y_1(t-1)].
\end{equation*}

If the current $\pi^*(t)=1$ and $\pi^{(m)}=0$, we can use the same method to prove (\ref{lemma_inequality}).

\section{Proof of Proposition 2}
Based on the conclusion in Proposition 1, $\ell_i^*(t)$ increases with $x_i(t)$ but $\ell^{(m)}(t)$ equals the constant $\ell_0(t)$. Thus, we only need to prove that myopic policy over-explores stochastic path $i$ (i.e., $\ell_i^*(t)\leq \ell^{(m)}(t)$) if $x_i(t)<\min\Big\{\frac{\alpha-\alpha_L}{\alpha_H-\alpha_L},\Bar{x}\Big\}$ and under-explores this path (i.e., $\ell_i^*(t)\geq \ell^{(m)}(t)$) if $x_i(t)>\max\Big\{\frac{\alpha-\alpha_L}{\alpha_H-\alpha_L},\Bar{x}\Big\}$. Then by the monotonicity of $\ell_i^*(t)$ in $x_i(t)$, we can prove the existence of $x^{th}$ to satisfy
\begin{equation}
    \min\Big\{\frac{\alpha-\alpha_L}{\alpha_H-\alpha_L},\Bar{x}\Big\} \leq x^{th}\leq \max\Big\{\frac{\alpha-\alpha_L}{\alpha_H-\alpha_L},\Bar{x}\Big\}.
\end{equation}

We first suppose that $\frac{\alpha-\alpha_L}{\alpha_H-\alpha_L}<\Bar{x}$. According to the definition of $\ell_i^*(t)$, we will prove $\ell_i^*(t)\leq \ell^{(m)}(t)$ for $x_i(t)=\frac{\alpha-\alpha_L}{\alpha_H-\alpha_L}$ by showing $Q_0^*(t+1)<Q_i^*(t+1)$, and then prove $\ell_i^*(t)\geq \ell^{(m)}(t)$ for $x_i(t)=\Bar{x}$ by showing $Q_0^*(t+1)>Q_i^*(t+1)$.

\subsection{Over-exploration Proof}
If $x_i(t)=\frac{\alpha-\alpha_L}{\alpha_H-\alpha_L}$ at current time, we have $\mathbb{E}[\alpha_i(t)|x_i(t)]=\alpha$. As the routing decision $\pi(t)=i$, the travel latency $\mathbb{E}[\ell_i(t+1)|x_i(t)=\frac{\alpha-\alpha_L}{\alpha_H-\alpha_L},y_i(t)]$ of this path $i$ at the next time slot is
\begin{align}
    &\mathbb{E}\Big[\ell_i(t+1)\Big|\frac{\alpha-\alpha_L}{\alpha_H-\alpha_L},y_i(t)\Big]\notag\\=&\mathbb{E}[\alpha_i(t)|x_i(t)]\mathbb{E}[\ell_i(t)|x_i(t-1),y_i(t-1)]+\Delta \ell.\label{E_l_t+1}
\end{align}
Similarly, if $\pi(t)=0$, the travel latency $\ell_{0}(t+1)$ of this deterministic path 0 at the next time slot is
\begin{equation*}
    \ell_0(t+1)=\alpha\ell_0(t)+\Delta \ell,
\end{equation*}
which equals $\mathbb{E}\big[\ell_i(t+1)\big|\frac{\alpha-\alpha_L}{\alpha_H-\alpha_L},y_i(t)\big]$ in (\ref{E_l_t+1}) under the condition that $\mathbb{E}[\alpha_i(t)|x_i(t)]=\alpha$. For any stochastic path $j$ with $j\neq i$, their expected travel latencies are not dependent on $\pi(t)=0$ or $\pi(t)=i$. However, as $x_i(t)<\Bar{x}$, we have $\mathbb{E}[x_i(t+1)]>x_i(t)$, making the future expected travel latency on path $i$ larger than the travel latency on path 0 under $\pi(t)=0$. In consequence, we have $Q_0^*(t+1)<Q_i^*(t+1)$, such that users over-explore path $i$ given $x_i(t)=\frac{\alpha-\alpha_L}{\alpha_H-\alpha_L}$.

\subsection{Under-exploration Proof}
If $x_i(t)=\Bar{x}$ at current time, we prove that $Q_0^*(t+1)> Q_i^*(t+1)$ is always true for any time horizon $T$ using mathematical induction. 

If $T=1$ and $\ell_0(t)<\mathbb{E}[\ell_i(t)|x_i(t-1),y_i(t-1)]$, myopic policy must choose path 0 but socially optimal policy may choose $i$. We have 
\[
\begin{aligned}
    Q_i^*(t+1)&\leq \mathbb{E}[\ell_0(t+1)|\pi(t)=i]\\&<\min_j \mathbb{E}[\ell_j(t+1)|\pi(t)=0]=Q^*_0(t+1),
\end{aligned}
\]
because $\mathbb{E}[\alpha_i(t)|x_i(t)=\Bar{x}]>\alpha$. 

We suppose $Q_0^*(t+1)> Q_i^*(t+1)$ is also true for time horizon $2\leq T\leq n$. Then we prove it is still true for time horizon $T=n+1$. Let $\pi_0^*(\tau)$ and $\pi_i^*(\tau)$ denote the two optimal policies for $Q_0^*(t+1)$ and $Q_i^*(t+1)$ after $t$, where $\tau\in\{t+1,\cdots,T+1\}$. If $\pi_0^*(\tau)=i$ again after time $t$, we let $\pi_i^*(\tau)=0$ to reach a lower cost $\mathbb{E}[\ell_{0}(\tau)|\pi(t)=i]< \mathbb{E}[\ell_i(\tau)|\pi(t)=0]$, due to the fact that $x_i(\tau)=\Bar{x}>\alpha$ for any $\tau$. Then for any other time slot $\tau\in\{t+1,\cdots,n+1\}$, we let $\pi_i^*(\tau)=\pi_0^*(\tau)$ to make $\mathbb{E}[\ell_{\pi_i^*(\tau)}(\tau)|\pi(t)=i]\leq \mathbb{E}[\ell_{\pi_{0}^*(\tau)}(\tau)|\pi(t)=0]$. After summing up all these costs, we can obtain $Q_0^*(t+1)> Q_i^*(t+1)$.

Then if $\frac{\alpha-\alpha_L}{\alpha_H-\alpha_L}>\Bar{x}$, we can use the similar methods as above to prove that $\ell_i^*(t)\geq \ell^{(m)}(t)$ for $x_i(t)=\Bar{x}$ by showing $Q_0^*(t+1)>Q_i^*(t+1)$, and then prove $\ell_i^*(t)\leq \ell^{(m)}(t)$ for $x_i(t)=\frac{\alpha-\alpha_L}{\alpha_H-\alpha_L}$ by showing $Q_0^*(t+1)<Q_i^*(t+1)$. And this completes the proof.

\section{Proof of Proposition 3}
When the discount factor $\rho=0$, the optimal policy is the same as myopic policy to only focus on the current cost. Thus, $\text{PoA}^{(m)}=\frac{1}{1-\rho}=1$ and the proposition holds. We next assume that $\rho\in(0,1)$ to show that $\text{PoA}^{(m)}\geq \frac{1}{1-\rho}$.

We consider the simplest two-path transportation network with $N=1$ to purposely choose the proper parameters to create the worst case. Let $\ell_0(t)=\frac{\Delta \ell}{1-\alpha}, x_1(t)=\Bar{x}, \mathbb{E}[\alpha_1(t)|\Bar{x}]=1,\ell_0(t)=\mathbb{E}[\ell_1(t)|\Bar{x},\emptyset],\alpha_L=0,q_{LL}\rightarrow 1$ and $\frac{\ell_0(t)}{\Delta \ell}\rightarrow\infty$. For other parameters, e.g., $q_{HH},\alpha_H$ and so forth, we can find proper values to satisfy the above constraints. Next we will calculate the social costs under the myopic policy and socially optimal policy, respectively.

\subsection{Social Cost under Myopic Policy}
We first calculate the the social cost under myopic policy. As $\ell_0(t)=\ell_1(t)$, the current user will myopically choose the safe path 0. Given $\ell_0(t)=\frac{\Delta\ell}{1-\alpha}$, we can obtain travel latency $\ell_0(t+1)$ as
\begin{align*}
    \ell_0(t+1)&=\alpha\ell_0(t)+\Delta\ell\\
    &=\frac{\Delta \ell}{1-\alpha}=\ell_0(t).
\end{align*}
This means even though users keep choosing this safe path 0, the travel latency on this path always equals $\frac{\Delta \ell}{1-\alpha}$.

As $\mathbb{E}[\alpha_1(t)|\Bar{x}]=1$ and $\ell_0(t)=\mathbb{E}[\ell_1(t)|\Bar{x},\emptyset]$ given $x_1(t-1)=\Bar{x}$, $y_1(t-1)=\emptyset$, we can obtain travel latency $\mathbb{E}[\ell_1(t+1)|\Bar{x},\emptyset]$ without exploration as
\begin{align*}
    \mathbb{E}[\ell_1(t+1)|\Bar{x},\emptyset]&=\mathbb{E}[\alpha_1(t)|x_1(t)=\Bar{x}] \mathbb{E}[\ell_1(t)|\Bar{x},\emptyset]\\
    &=\mathbb{E}[\ell_1(t)|\Bar{x},\emptyset].
\end{align*}
Since $x_1(t)=\Bar{x}$, the expectation $\mathbb{E}[x_1(t+1)]=\Bar{x}$. Thus, the travel latency on this path also keeps at $\mathbb{E}[\ell_1(t)|\Bar{x},\emptyset]$ without any user exploration.

Under these parameters, myopic users keep choosing safe path 0 and will never explore risky path 1. Thus, we calculate the corresponding social cost as
\begin{align*}
    C^{(m)}(\mathbf{L}(t),\mathbf{x}(t))&=\ell_0(t)+\rho \ell_0(t) +\cdots\\
    &=\frac{\ell_0(t)}{1-\rho}.
\end{align*}

\subsection{Minimum Social Cost}
Next, we will calculate the social cost under socially optimal policy. We let the current user explore path 1 to exploit the possible $\alpha_L=0$. As $q_{LL}\rightarrow 1$, $q_{HH}\rightarrow 1$, $p_H\rightarrow1$ and $p_L\rightarrow 0$, we can obtain that $P(y_1(t)=1)\rightarrow 0$ and $P(y_1(t)=0)\rightarrow 1$ as the system has been running for a long time. Then we have
\begin{align*}
    C^*(\mathbf{L}(t),\mathbf{x}(t))\leq& \mathbb{E}[\ell_1(t)|\Bar{x},\emptyset]\\&+\text{Pr}(y_1(t)=1)\sum_{k=1}^\infty\rho^k\ell_0(t+k)\\&+\text{Pr}(y_1(t)=0)\sum_{m=1}^\infty\rho^m \mathbb{E}[\ell_1(t+m)|0,0],
\end{align*}
where $x_1(t+m)\rightarrow 0$ and $y_1(t+m)=0$ given $p_L\rightarrow 0$. The above inequality tells that if $y_1(t)=0$ with probability $\text{Pr}(y_1(t)=0)$, the socially optimal policy let latter users keep exploit the low travel latency on path 0, while if $y_1(t)=1$ with probability $\text{Pr}(y_1(t)=1)$, the socially optimal policy let the latter users go back to path 0 again. We can further obtain that 
\begin{align*}
    C^*(\mathbf{L}(t),\mathbf{x}(t))&\leq \mathbb{E}[\ell_1(t)|\Bar{x},\emptyset]+\frac{\rho}{1-\rho}\Delta\ell\\
    &=\ell_0(t)+\frac{\rho}{1-\rho}\Delta\ell.
\end{align*}
Finally, we can obtain 
\begin{align*}
    \text{PoA}^{(m)}&=\frac{C^{(m)}(\mathbf{L}(t),\mathbf{x}(t))}{C^*(\mathbf{L}(t),\mathbf{x}(t))}\\&\geq \frac{1}{1-\rho},
\end{align*}
by letting $\frac{\Delta\ell}{\ell_0(t)}\rightarrow 0$.

\section{Proof of Proposition 4}
To tell the huge PoA result and selfish users' deviation from the optimal routing recommendations $\pi^*(t)$, we consider the simplest two-path transportation network. Initially, the safe path 0 has $\ell_0(t=0)=0$ with $\alpha\rightarrow 1$ and the risky path 1 has an arbitrarily travel latency $\ell_1(0)$. We let $\Bar{x}=0$ and $\mathbb{E}[\alpha_1(t)|\Bar{x}]=0$ by setting $q_{LL}=1$ and $q_{HH}\neq 1$. Then selfish users always choose $\pi^{\emptyset}(t)=1$ from $t=0$. We calculate the social cost
\begin{align*}
    C^{\emptyset}(\ell_0(0),\ell_1(0),\Bar{x})&=\ell_1(0)+\sum_{j=1}^\infty \rho^j \Delta \ell\\
    &=\ell_1(0)+\frac{\rho \Delta \ell}{1-\rho},
\end{align*}
based on the fact that $\mathbb{E}[\alpha_1(t)|\Bar{x}]=0$ and $\mathbb{E}[\ell_1(t)|\Bar{x},y_1(t-1)]=\Delta\ell$ for any $t\geq 1$. 

However, socially optimal policy let users to path 0 to exploit the small travel latency $\ell_0(0)$. In this case, we have
\begin{align*}
    C^*(\ell_0(0),\ell_1(0),\Bar{x})&=\ell_0(0)+\sum_{j=2}^\infty\rho^j\Delta\ell\\
    &=\frac{\rho^2\Delta\ell}{1-\rho},
\end{align*}
where $\ell_0(0)=0$, $\mathbb{E}[\ell_1(1)|\Bar{x},\emptyset]=0$ and $\mathbb{E}[\ell_1(t)|\Bar{x},y(t-1)]=\Delta\ell$ for ant $t\geq 2$.

In this case, we obtain 
\begin{align*}
    \text{PoA}^{\emptyset}&=\max_{\ell_0(0),\ell_1(0),\Bar{x},\rho,\Delta\ell} \frac{C^{\emptyset}(\ell_0(0),\ell_1(0),\Bar{x})}{ C^*(\ell_0(0),\ell_1(0),\Bar{x})}\\&=\frac{(1-\rho)\ell_1(0)}{\rho^2\Delta\ell}+\rho\\ &=\infty,
\end{align*}
where we let $\frac{(1-\rho)\ell_1(0)}{\rho^2\Delta\ell}\rightarrow\infty$.

Next, we analyze that even if the mechanism provides the optimal recommendation $\pi^*(t)$, the PoA is still infinite. Given the same belief on the initial travel latency $\ell_0(0)$ and $\ell_1(0)$, the selfish user will always choose the path $i$ because
\begin{align*}
    \text{Pr}\Big(\mathbb{E}[\ell_0(t)]>\mathbb{E}[\ell_i(t)|\Bar{x},y_i(t-1)]\Big)\rightarrow 1
\end{align*}
given $\mathbb{E}[\alpha_i(t)|\Bar{x}]=0$ and $\alpha\rightarrow 1$. Thus, the information hiding mechanism may not work given $\Bar{x}<\frac{\alpha-\alpha_L}{\alpha_H-\alpha_L}$, and it still makes $\text{PoA}^{\emptyset}=\infty$.

\section{Proof of Theorem 1}
We first prove that selfish users will follow our mechanism's optimal routing recommendations if $\Bar{x}>\frac{\alpha-\alpha_L}{\alpha_H-\alpha_L}$. After that, we prove that the worst-case PoA is reduced to $\frac{1}{1-\frac{\rho}{2}}$ under $\Bar{x}<\frac{\alpha-\alpha_L}{\alpha_H-\alpha_L}$.

\subsection{Proof of SID Mechanism's Efficiency}
Note that if $\mathbb{E}[\alpha_i(t)|\Bar{x}]>1$, the expected travel latency in path $i$ keeps increasing exponentially in $\mathbb{E}[\alpha_i(t)|\Bar{x}]$. Given the system has been running for a long time, the socially optimal policy will never choose this path, either. Thus, we only consider the more practical case $\mathbb{E}[\alpha_i(t)|\Bar{x}]\leq 1$ in the following. 

Lacking any historical information of the hazard belief and assuming that the mechanism operates already a long time, the current user's best estimate of the travel latency of path $i$ is the stationary distribution $P^*(\Bar{\ell}_i(t))$ of $\Bar{\ell}_i(t)$ under optimal policy $\pi^*(t)$. We do not need to obtain $P^*(\Bar{\ell}_i(t))$ but can use it to estimate the long-run average un-discounted travel latency $\lambda^*$ of all path:
\begin{align*}
    \lambda^*&=\int_{A}\Bar{\ell}_i(t)dP^*(\Bar{\ell}_i(t))+\int_{B}\Bar{\ell_0}dP^*(\Bar{\ell}_i(t))\\
    &\leq \Bar{\ell}_0,
\end{align*}
where socially optimal policy chooses path $\pi^*(t)=i$ when $\Bar{\ell}_i(t)$ is in region A and chooses path $\pi^*(t)=0$ when $\Bar{\ell}_i(t)$ is in region B.

If the platform's recommendation is $\pi^*(t)=0$ for the current user, he will follow this recommendation to path 0. Otherwise, he will calculate
\begin{align*}
    &\mathbb{E}_{P^*(\Bar{\ell}_i(t))}[\Bar{\ell}_i(t)|\Bar{\ell}_i(t)\in A]\\=&\frac{\int_{A}\Bar{\ell}_i(t)dP^*(\Bar{\ell}_i(t))}{\int_{A}dP^*(\Bar{\ell}_i(t))}\\ =&\frac{\lambda^*-\int_{B}\Bar{\ell_0}dP^*(\Bar{\ell}_i(t))}{\int_{A}dP^*(\Bar{\ell}_i(t))}\\ \leq&\frac{\Bar{\ell}_0-\int_{B}\Bar{\ell_0}dP^*(\Bar{\ell}_i(t))}{\int_{A}dP^*(\Bar{\ell}_i(t))}=\Bar{\ell}_0(t).
\end{align*}
Thus, each user will follow the optimal recommendation to choose path $\pi^*(t)=i$ given $\Bar{x}>\frac{\alpha-\alpha_L}{\alpha_H-\alpha_L}$. This is because the former exploration to risky path $i$ may learn a $\alpha_L$ there to greatly reduce $\mathbb{E}[\ell_i(t)|x_i(t-1),y_i(t-1)=0]$.

\subsection{Proof of $\text{PoA}^{(\text{SID})}\leq \frac{1}{1-\frac{\rho}{2}}$}
Based on the former analysis, we can further prove $\text{PoA}^{(\text{SID})}\leq \frac{1}{1-\frac{\rho}{2}}$, which only happens when $\Bar{x}<\frac{\alpha-\alpha_L}{\alpha_H-\alpha_L}$. With the information disclosure $\mathbf{L}(t)$, selfish users will deviate to follow myopic policy $\pi^{(m)}(t)$. We consider the maximum over-exploration in the simplest two-path network to show the bounded PoA.

Let $\ell_0(0)=\ell_1(0)-\varepsilon$ for path 0 with $\alpha\rightarrow 1$ to keep the travel latency on path 0 unchanged without user routing, where $\varepsilon$ is positive infinitesimal. We set $\ell_1(0)=\frac{\Delta \ell}{1-\mathbb{E}[\alpha_1(0)|\Bar{x}]}$ for stochastic path 1 with $x_1(0)=\Bar{x}$, such that the expected travel latency at the next time $t=1$ is
\begin{align*}
    \mathbb{E}_{y_1(0)}[\ell_1(1)|\Bar{x},y_1(0)]&=\mathbb{E}[\alpha_1(0)|\Bar{x}]\ell_1(0)+\Delta \ell \\ &=\frac{\Delta\ell}{1-\mathbb{E}[\alpha_1(0)|\Bar{x}]},
\end{align*}
which keeps as $\ell_1(0)$ all the time with users' continuous explorations.
Then users keep choosing this risky path 1 without exploitation to path 0, and we can calculate the social cost
\begin{align*}
    C^{(m)}(\ell_0(0),\ell_1(0),\Bar{x})&=\sum_{j=0}^\infty\rho^j\ell_0(0)
    \\&=\frac{\ell_1(0)}{1-\rho}.
\end{align*}
However, the socially optimal policy makes $\pi^*(0)=0$ for the first user to bear the similar travel latency on path 0. Then the expected travel latency on the next time slot on path 1 is reduced to \[\mathbb{E}[\ell_1(1)|\Bar{x},\emptyset]=\mathbb{E}[\alpha_1(0)|\Bar{x}]\ell_1(0).\]
After the first exploitation on path 0, the travel latency on this path is increases to $\ell_0(0)+\Delta\ell$. While the expected travel latency on path 1 is always less than $\ell_1(0)$ because
\begin{align*}
    \mathbb{E}[\ell_1(t+1)|\Bar{x},y_1(t)]=&\Bar{\alpha}_1\mathbb{E}[\ell_1(t)|\Bar{x},y_1(t-1)]+\Delta\ell\\ < &\mathbb{E}[\alpha_1(0)|\Bar{x}]\ell_1(0)+\Delta\ell=\ell_1(0)
\end{align*}
for any time $t\geq 1$. 

Note that if $\mathbb{E}[\alpha_1(t)|\Bar{x}]=0$, then the expected travel latency on path 1 is always $\Delta\ell$, and socially optimal policy will not choose path 0, either. If $\mathbb{E}[\alpha_1(t)|\Bar{x}]=1$, then the expected travel latencies on both paths are infinite with $\frac{\Delta\ell}{1-\mathbb{E}[\alpha_1(0)|\Bar{x}]}\rightarrow\infty$, and the $\text{PoA}^{\emptyset}=1$. Hence, the worst-case does not happen when $\mathbb{E}[\alpha_1(t)|\Bar{x}]=0$ or $1$. We next derive the $\mathbb{E}[\alpha_1(0)|\Bar{x}]$ in the worst-case, which is denoted by $\Bar{\alpha}$. From the evolution of $\mathbb{E}[\ell_1(t+1)|\Bar{x},y_1(t)]$, we aim to minimize the first order derivative below
\begin{align*}
    &\frac{\partial (\mathbb{E}[\ell_1(t+1)|\Bar{x},y_1(t)]-\mathbb{E}[\ell_1(t)|\Bar{x},y_1(t-1)])}{\partial \mathbb{E}[\ell_1(t-2)|\Bar{x},y_1(t-2)]}\\
    =&\frac{\Delta\ell +(\Bar{\alpha}_1-1)(\Bar{\alpha}_1\mathbb{E}[\ell_1(t-2)|\Bar{x},y_1(t-2)]+\Delta \ell)}{\partial \mathbb{E}[\ell_1(t-2)|\Bar{x},y_1(t-2)]}\\
    =&\Bar{\alpha}_1^2-\Bar{\alpha}_1,
\end{align*}
where the minimum is reached at $\Bar{\alpha}_1=\frac{1}{2}$. Note that $\Bar{\alpha}_1=\frac{1}{2}$ well balances the expected travel latency and a single user's incurred latency $\Delta\ell$ to make the largest PoA. 

Then we can calculate the optimal social cost as
\begin{align*}
    C^*(\ell_0(0),\ell_1(0),\Bar{x})=& \ell_0(0)+\rho\Bar{\alpha}_1\ell_1(0)\\&+\rho^2(\Bar{\alpha}^2_1\ell_1(0)+\Delta \ell)+\cdots\\
    \geq &\ell_0(0)+ \sum_{j=1}^{\infty}\rho\frac{\ell_1(0)}{2}\\=&\ell_0(0)+\frac{\rho \ell_1(0)}{2-2\rho}.
\end{align*}
Though socially optimal policy may still choose path 0 to reduce the expected latency for path 1 after a period, the caused average expected travel latency is still no less than $\frac{\ell_1(0)}{2}$. This is because the travel latency on path 0 increases to $\ell_0(0)+\Delta \ell$ after the first exploitation. 

Finally, we can obtain the worst case PoA as
\begin{align*}
    \text{PoA}^{(\text{SID})}&=\max \frac{C^{(m)}(\mathbf{L}(t),\mathbf{x}(t)}{C^*(\mathbf{L}(t),\mathbf{x}(t))}\\&\leq \frac{\frac{\ell_1(0)}{1-\rho}}{\ell_0(0)+\frac{\rho \ell_1(0)}{2-2\rho}}\\&=\frac{1}{1-\frac{\rho}{2}}.
\end{align*}
This completes the proof.

\end{document}